\documentclass{article}
\usepackage{amsmath, amssymb, amsfonts}
\title{ Thin shell embedded in de Sitter space} 
\author{Hristu Culetu, \\Ovidius University, Department of Physics and Electronics, \\ Mamaia Avenue 124, 900527 Constanta, Romania, \\e-mail : hculetu@yahoo.com}

\begin{document}
\numberwithin{equation}{section}
\pagenumbering{arabic}
\maketitle
\newcommand{\fv}{\boldsymbol{f}}
\newcommand{\tv}{\boldsymbol{t}}
\newcommand{\gv}{\boldsymbol{g}}
\newcommand{\OV}{\boldsymbol{O}}
\newcommand{\wv}{\boldsymbol{w}}
\newcommand{\WV}{\boldsymbol{W}}
\newcommand{\NV}{\boldsymbol{N}}
\newcommand{\hv}{\boldsymbol{h}}
\newcommand{\yv}{\boldsymbol{y}}
\newcommand{\RE}{\textrm{Re}}
\newcommand{\IM}{\textrm{Im}}
\newcommand{\rot}{\textrm{rot}}
\newcommand{\dv}{\boldsymbol{d}}
\newcommand{\grad}{\textrm{grad}}
\newcommand{\Tr}{\textrm{Tr}}
\newcommand{\ua}{\uparrow}
\newcommand{\da}{\downarrow}
\newcommand{\ct}{\textrm{const}}
\newcommand{\xv}{\boldsymbol{x}}
\newcommand{\mv}{\boldsymbol{m}}
\newcommand{\rv}{\boldsymbol{r}}
\newcommand{\kv}{\boldsymbol{k}}
\newcommand{\VE}{\boldsymbol{V}}
\newcommand{\sv}{\boldsymbol{s}}
\newcommand{\RV}{\boldsymbol{R}}
\newcommand{\pv}{\boldsymbol{p}}
\newcommand{\PV}{\boldsymbol{P}}
\newcommand{\EV}{\boldsymbol{E}}
\newcommand{\DV}{\boldsymbol{D}}
\newcommand{\BV}{\boldsymbol{B}}
\newcommand{\HV}{\boldsymbol{H}}
\newcommand{\MV}{\boldsymbol{M}}
\newcommand{\be}{\begin{equation}}
\newcommand{\ee}{\end{equation}}
\newcommand{\ba}{\begin{eqnarray}}
\newcommand{\ea}{\end{eqnarray}}
\newcommand{\bq}{\begin{eqnarray*}}
\newcommand{\eq}{\end{eqnarray*}}
\newcommand{\pa}{\partial}
\newcommand{\f}{\frac}
\newcommand{\FV}{\boldsymbol{F}}
\newcommand{\ve}{\boldsymbol{v}}
\newcommand{\AV}{\boldsymbol{A}}
\newcommand{\jv}{\boldsymbol{j}}
\newcommand{\LV}{\boldsymbol{L}}
\newcommand{\SV}{\boldsymbol{S}}
\newcommand{\av}{\boldsymbol{a}}
\newcommand{\qv}{\boldsymbol{q}}
\newcommand{\QV}{\boldsymbol{Q}}
\newcommand{\ev}{\boldsymbol{e}}
\newcommand{\uv}{\boldsymbol{u}}
\newcommand{\KV}{\boldsymbol{K}}
\newcommand{\ro}{\boldsymbol{\rho}}
\newcommand{\si}{\boldsymbol{\sigma}}
\newcommand{\thv}{\boldsymbol{\theta}}
\newcommand{\bv}{\boldsymbol{b}}
\newcommand{\JV}{\boldsymbol{J}}
\newcommand{\nv}{\boldsymbol{n}}
\newcommand{\lv}{\boldsymbol{l}}
\newcommand{\om}{\boldsymbol{\omega}}
\newcommand{\Om}{\boldsymbol{\Omega}}
\newcommand{\Piv}{\boldsymbol{\Pi}}
\newcommand{\UV}{\boldsymbol{U}}
\newcommand{\iv}{\boldsymbol{i}}
\newcommand{\nuv}{\boldsymbol{\nu}}
\newcommand{\muv}{\boldsymbol{\mu}}
\newcommand{\lm}{\boldsymbol{\lambda}}
\newcommand{\Lm}{\boldsymbol{\Lambda}}
\newcommand{\opsi}{\overline{\psi}}
\renewcommand{\tan}{\textrm{tg}}
\renewcommand{\cot}{\textrm{ctg}}
\renewcommand{\sinh}{\textrm{sh}}
\renewcommand{\cosh}{\textrm{ch}}
\renewcommand{\tanh}{\textrm{th}}
\renewcommand{\coth}{\textrm{cth}}

\begin{abstract}
The behaviour of a static thin shell embedded in dS space is investigated. To satisfy the junction conditions at the boundary, one rescales the time variable. The surface energy $\sigma$ on the shell is positive but its surface tension $\tau$ is negative when $r << a$, where $a$ is the horizon radius. The repulsive character of the radial acceleration of a static inner observer is fortified by the process of sinking in the Sitter universe. An upper limit for $\sigma$ has to be imposed for the shell mass to remain positive.
 \end{abstract}
 
\section{Introduction}
 To find whether the cosmological expansion takes place locally (say, at the Solar System level) is a very complicated matter and is still an unsolved problem \cite{TP}. The solution depends on the model of the Universe. The purpose is to combine both classes of solutions (cosmological and local) and to find out exact solutions for the gravitational field of a particular system immersed in a cosmological background, albeit a simple superposition of solutions is not in total agreement with the nonlinear feature of the theory \cite{MV, CG, FJ, MSFX}.

  As Padmanabhan has shown \cite{TP}, even the notion of ''expansion'' is relativeand depends on the frame we choose; for example, the Schwarzschild metric (S) is expanding for a geodesic observer. We will work in the ''thin shell'' approximation, namely the thickness of the shell is small compared to any length scale from the physical system \cite{BGG}. 
	
	The theory of surface layers in General Relativity (GR) was developed by Israel \cite{WI}, making use of the Gauss-Codazzi equations (see also \cite{BGG, CL, IS, KKP, HC, MH}). The first Israel junction condition states that the induced metric on the spherical shell (bubble) should be continuous. We will next use the second junction condition relating the jump of the extrinsic curvature across the static shell to the stress tensor on the shell. 
	
	Our motivation in this paper is to look for the connections between the quantities related to the intrinsic geometry of the shell (the surface energy and surface tension) and geometric properties of the two spacetimes from either side of it. In Sec. 2 we present briefly the well-known case of a thin shell embedded in Minkowski space and write down the equations relating the shell mass to the intrinsic parameters of the hypersurface. Sec. 3 studies the more complex situation when the shell is immersed in de Sitter spacetime. The static character of the system is investigated in terms of the physical quantities characterising the shell and the inner and outer spacetimes.
	
	Throughout the paper we use geometrical units $G = c = 1$, unless otherwise specified.

	\section{ Static thin shell}
	The next step is to establish the geometries on either side of the spherical shell of mass $m$, measured by a static observer on its surface. Outside it the metric is Schwarzschild, whether the shell is static or not
  \begin{equation}
   ds_{+}^{2} = -\left(1 - \frac{2m}{r}\right) dt^{2} + \frac{1}{1 - \frac{2m}{r}} dr^{2} + r^{2} d \Omega^{2},     
 \label{2.1}
 \end{equation} 
	where $d \Omega^{2}$ stands for the metric on the unit 2-sphere. Birkhoff's theorem guarantees that the interior geometry is flat
	  \begin{equation}
   ds_{-}^{2} = -\left(1 - \frac{2m}{R}\right) dt^{2} + dr^{2} + r^{2} d \Omega^{2},     
 \label{2.2}
 \end{equation} 
with $R$ - the (constant) radius of the shell, $R > 2m$. We chose the above form of the inner metric for to work with the same time variable on both regions (a choice that results from the first junction condition at $r = R$). 

 Let us take $S_{ab}$ as the perfect fluid source with $\sigma$ - the surface energy density and $\tau$ - the surface tension ($\tau = - p$, where $p$ is the surface pressure)
     \begin{equation}
		S_{ab} = (\sigma - \tau) u_{a}u_{b} - \tau h_{ab},
 \label{2.3}
 \end{equation} 
where $u^{a}$ is the velocity vector field ($u^{a}u_{a} = -1$), $h_{ab} = g_{ab} - \xi_{a} \xi_{b}$, with $\xi^{a}$ - the unit normal vector to the surface $r = R$, $\xi^{a}u_{a} = 0, ~ \xi^{a} \xi_{a} = 1$ ($a, b$ above span the coordinates (t, r, $\theta, \phi$)). 

The stress tensor $S_{ab}$ is related to the jump of the extrinsic curvature $[K_{ab}] = K_{ab}^{+} - K_{ab}^{-}$ (here $a,b  = t, \theta, \phi$) through the Lanczos equation \cite{BGG, CL, WI}
     \begin{equation}
		[K_{ab}] -  h_{ab} [K^{c}_{~c}] = -8\pi S_{ab}.
 \label{2.4}
 \end{equation} 
To compute the components of $K_{ab}$ we make use of the Kolekar-Kothawala-Padmanabhan formula \cite{KKP} 
     \begin{equation}
		K_{ab} = -\frac{\sqrt{g} f'}{2f} u_{a}u_{b} + \frac{\sqrt{g}}{r} q_{ab},
 \label{2.5}
 \end{equation} 
where $f(r) = -g_{tt},~g(r) = 1/g_{rr}~, f' = df/dr$ and $q_{ab} = h_{ab} + u_{a} u_{b}$. The detailed calculations appear in many textbooks (even for a dynamical situation) so we give here the main results for our particular case
     \begin{equation}
		\sigma = \frac{m}{2\pi R^{2} (1 + \sqrt{1 - \frac{2m}{R}})},~~~\tau = \frac{\sqrt{1 - \frac{2m}{R}} + \frac{m}{R} - 1}{8\pi R \sqrt{1 - \frac{2m}{R}}}
 \label{2.6}
 \end{equation} 
and
 \begin{equation}
	m = 4\pi \sigma R^{2} (1 - 2\pi \sigma R).	
 \label{2.7}
 \end{equation} 
It is worth noticing that we have always $\tau < 0$ and $\sigma > 0$. The direct relation between them is
     \begin{equation}
		\tau(\sigma) = -\frac{\pi R \sigma^{2}}{1 - 4\pi \sigma R},
 \label{2.8}
 \end{equation} 
where we have $1 - 4\pi \sigma R = \sqrt{1 - \frac{2m}{R}} > 0$ (obtained from (2.7)), or $\sigma < 1/4\pi R$. This last condition confirms that $m$ is always positive.

 Let us observe that when $R >> m$, $\tau \approx 0$ and $\sigma = m/4\pi R^{2}$, i.e. the classical situation ($\tau$ is negligible and $\sigma$ = energy/area). In contrast, when $R \rightarrow 2m$, we have $\sigma = 1/8\pi m$ and the surface tension $\tau \rightarrow -\infty$, a consequence of the divergence of the proper S acceleration on the horizon $r = R = 2m$. We have to keep in mind that the equation for $\tau$ originates from the discontinuity of the normal components of the acceleration when the shell is crossed \cite{IS, CL}.

\section{Thin shell embedded in de Sitter space}
We pass now to a cosmological situation and consider that our shell is immersed in a dS space, so that the outer metric becomes SdS
  \begin{equation}
   ds_{+}^{2} = -\left(1 - \frac{2m}{r} - \frac{r^{2}}{a^{2}}\right) dt^{2} + \frac{1}{1 - \frac{2m}{r} - \frac{r^{2}}{a^{2}}} dr^{2} + r^{2} d \Omega^{2}, ~~~R \leq r < a,    
 \label{3.1}
 \end{equation} 
where the constant $a$ is the horizon radius of the dS space, with $a = \sqrt{3/\Lambda},~\Lambda$ being the cosmological constant. We take $a > 3\sqrt{3}m$, hence (3.1) preserves its static form \cite{SS}. The interior geometry will be considered to have the form
  \begin{equation}
   ds_{-}^{2} = -\left(1 - \frac{2m}{R} - \frac{r^{2}}{a^{2}}\right) dt^{2} + \frac{1}{1 - \frac{r^{2}}{a^{2}}} dr^{2} + r^{2} d \Omega^{2}, ~~~ r \leq R ,    
 \label{3.2}
 \end{equation} 
so that the Minkowski space is retrieved when $a \rightarrow \infty$. In addition, we suppose $r < a \sqrt{1 - \frac{2m}{R}}$, for any $r < R$, for to avoid a signature flip of the $g_{tt}$ metric coefficient. That condition is reasonable because the radius $a$ is of cosmological order, namely $r << a$. The shell radius should not, of course, be too close to $2m$. 

The curved metric (3.2) is generated by the stress tensor $T_{ab}$, obtained from Einstein's equations $G_{ab} = 8\pi T_{ab}$. It consists of an anisotropic fluid to which the energy density $\rho$, the radial pressure $p_{r}$ and tangential pressures $p_{\theta}$ and $p_{\phi}$ are given by
  \begin{equation}
	\rho = \frac{3}{8\pi a^{2}},~~~p_{r} = \rho - \frac{m }{2\pi Ra^{2} \left(1 - \frac{2m}{R} - \frac{r^{2}}{a^{2}}\right)}
 \label{3.3}
 \end{equation} 
and
  \begin{equation}
	   p_{\theta} = p_{\phi} = p_{r} - \frac{m r^{2}}{4\pi Ra^{4} \left(1 - \frac{2m}{R} - \frac{r^{2}}{a^{2}}\right)^{2}}.
 \label{3.4}
 \end{equation} 
We notice that $\rho$ is constant and positive and $\rho, p_{r}$ and $p_{\theta}$ are vanishing when $a \rightarrow \infty$ (the interior geometry becomes flat). We also observe that 
$p_{r} < 0$ and $p_{\theta} < 0$ for any values of $r < R,~R > 2m$ and $R << a$. Moreover, $\rho < |p_{r}|$ and the energy conditions are not observed. That is a consequence of the negative pressures of the dS spacetime. When $m = 0$ (shell removed), one obtains  $\rho = -p_{r} = -p_{\theta} = 3/8\pi a^{2}$, as it should be. In addition, $p_{r}, p_{\theta}$ are monotonically descending functions of $r$ for $0 < r < R$. 

 Let us take now a congruence of static timelike observers given by the velocity vector field $u^{a} = (1/\sqrt{1 - \frac{2m}{R} - \frac{r^{2}}{a^{2}}}, 0, 0, 0)$. The covariant acceleration $a^{b} = u^{a}\nabla_{a}u^{b}$ has only one nonzero component
  \begin{equation}
a^{r}_{-} = - \frac{r}{a^{2}} - \frac{2mr }{Ra^{2} \left(1 - \frac{2m}{R} - \frac{r^{2}}{a^{2}}\right)}.	
 \label{3.5}
 \end{equation} 
It is clear from (3.5) that the effect of $m$ on $a^{r}$ is to strenghten its negativity. This is in contrast with the radial acceleration of a static observer located in the exterior space (3.1)
  \begin{equation}
a^{r}_{+} = - \frac{r}{a^{2}} + \frac{m }{r^{2}}. 
 \label{3.6}
 \end{equation} 
One sees that the second term on the r.h.s. of (3.6) weakens the value of $a^{r}_{+}$, so its repulsive character is diminished. The first term on the r.h.s. of (3.5) is the obvious dS  term but the second one depends on the shell parameters $m$ and $R$. Its repulsive feature is highlighted once $r$ is increased from zero  to $R$. 

 To study the behaviour of $a^{r}_{-}$ in the region $r << a$, we neglect $r^{2}/a^{2}$ w.r.t. unity at the denominator and get
  \begin{equation}
a^{r}_{-} = - \frac{r}{a^{2}} \frac{1}{1 - \frac{2m}{R}}.
 \label{3.7}
 \end{equation} 
If $R$ is close to $2m$, say $R = 3m$, one finds that $a^{r}_{-} = -3r/a^{2}$, which is three times the value in the absence of the shell. 

We investigate now the junction conditions at the boundary $r = R$ of the spacetimes (3.1) - (3.2). With the help of (2.3) - (2.5), the equations for $\sigma$ and $\tau$ appears now as
  \begin{equation}
	\sqrt{1 - \frac{R^{2}}{a^{2}}} - \sqrt{1 - \frac{2m}{R} - \frac{R^{2}}{a^{2}}} = 4\pi \sigma R
 \label{3.8}
 \end{equation} 
and
  \begin{equation}
	\frac{4\pi \sigma R^{2}}{a^{2} \left(1 - \frac{2m}{R} - \frac{R^{2}}{a^{2}}\right)} + \frac{m}{R^{2} \sqrt{1 - \frac{2m}{R} - \frac{R^{2}}{a^{2}}}} = 4\pi (\sigma - 2\tau)                           
 \label{3.9}
 \end{equation} 
and the shell mass is obtained from (3.8)
 \begin{equation}
	m = 4\pi \sigma R^{2} \left( \sqrt{1 - \frac{R^{2}}{a^{2}}} - 2\pi \sigma R\right).	
 \label{3.10}
 \end{equation} 
While $\sigma$ is given by (3.8) (in terms of the constant $m, a, R$), for the surface tension the above equations give us
  \begin{equation}
	8\pi R \tau = \frac{(1 - \frac{2m}{R} - \frac{2R^{2}}{a^{2}}) \sqrt{1 - \frac{R^{2}}{a^{2}}}}{1 - \frac{2m}{R} - \frac{R^{2}}{a^{2}}} - \frac{1 - \frac{m}{R} - \frac{2R^{2}}{a^{2}}}{\sqrt{1 - \frac{2m}{R} - \frac{R^{2}}{a^{2}}}}
 \label{3.11}
 \end{equation} 
One notices that the expressions (2.6) and (2.7) are recovered from (3.10) and (3.11) when $a \rightarrow \infty$.

\section{Conclusions}
To find out exact solutions of the Einstein equations for a particular system immersed in a cosmological background is a difficult task. That originates from the nonlinear character of the field equations in GR such that a linear superposition of solutions is not always appropriate. The static situation is simpler but, nevertheless, one needs a negative surface tension on the boundary for the matching conditions to be satisfied. An important consequence of the immersion of the shell in the dS space is the increase of the repulsive force (acting as expansion) on a static interior observer.

\end{document}